\documentclass[twocolumn,english,aps,superscriptaddress,prl]{revtex4-2}

\usepackage[colorlinks=true,allcolors=blue]{hyperref}
\usepackage{array}
\usepackage[latin9]{inputenc}
\usepackage{amsmath}
\usepackage{amssymb}
\usepackage{graphicx}
\usepackage{babel}
\usepackage{mathrsfs}
\usepackage{amsfonts}
\usepackage{epstopdf}
\usepackage{multirow}
\usepackage{color}
\usepackage{natbib}
\usepackage{bm}
\usepackage{amsthm}
\usepackage{siunitx}
\usepackage{quantikz}

\newcommand{\dg}{\dagger}

\newcommand{\br}{\mathbf{r}}

\newcommand{\ii}{\mathrm{i}}

\newcommand{\fbk}{\mathrm{FBK}}
\newcommand{\tchi}{\widetilde{\chi}}
\usepackage{braket}
\usepackage{xcolor}

\usepackage[normalem]{ulem}

\begin{document}

\title{Unleashing Emergent Fermions with Rydberg Atom Simulators}
\author{Hanteng Wang}
\affiliation{
Institute for Advanced Study, Tsinghua University, Beijing 100084, China
}

\author{Xingyu Li}
\affiliation{
Institute for Advanced Study, Tsinghua University, Beijing 100084, China
}

\author{Shang Liu}
\affiliation{Institute of Physics,
Chinese Academy of Sciences, Beijing 100910, China}

\author{Yingfei Gu}
\affiliation{
Institute for Advanced Study, Tsinghua University, Beijing 100084, China
}

\author{Chengshu Li}
\email{chengshu@mail.tsinghua.edu.cn}
\affiliation{
Institute for Advanced Study, Tsinghua University, Beijing 100084, China
}
\date{\today}

\begin{abstract}
Rydberg atom simulators, in both analog and digital modes, have attracted significant recent interest due to their versatile geometric reconfigurability. In this work, leveraging this feature, we propose two complementary approaches, one for each mode, to characterize emergent fermions in critical quantum many-body systems. In the analog mode, we assemble the Rydberg atoms in a ``developable'' (namely, preserving local couplings) M\"obius band geometry to realize antiperiodic boundary conditions, where fermionic states reside. Spectroscopic measurement in this sector then reveals universal energy ratios of the bosonic and fermionic states. In the digital mode, we carry out a fermionic version of Kibble--Zurek ramping with a quantum circuit, directly addressing the fermionic scaling form. Reconfigurability allows an exponential speed-up of this task, with an $\mathcal{O}(\log L\log\log L)$ circuit-depth overhead. Our work establishes the Rydberg atom simulator as a uniquely powerful platform to attack the notoriously difficult issue of experimentally probing emergent fermions that are nonlocally defined in a bosonic system.
\end{abstract}

\maketitle
Programmable Rydberg atom arrays have rapidly developed into a versatile platform for quantum many-body physics~\cite{browaeys20}, combining large-scale analog simulation with increasingly capable gate-based control~\cite{bernien2017probing,evered2023high,bluvstein2024logical}. A central advantage of this platform is its geometric reconfigurability: atoms can be arranged in flexible geometries, rearranged during a protocol, and coupled through high-fidelity entangling operations \cite{Barredo2018,Bluvstein2022}. These capabilities have opened new routes for investigating strongly correlated matter with exotic phases \cite{Ebadi2021,Scholl2021,Semeghini2021}, gauge structures \cite{surace2020lattice,cheng2024emergent,Bylinskii2025,lilin2025}, and critical phenomena \cite{keesling2019quantum,fang2024probing,zhang2025observation}, going beyond what rigid architectures can access.

In mainstream array experiments, the atoms are pinned to fixed sites by optical tweezers and prohibited from hopping. Hence, the underlying hardware is intrinsically bosonic. This raises a natural question: can such a bosonic platform directly reveal fermionic structures? Such structures can emerge through nonlocal Jordan--Wigner (JW)-type transformations~\cite{Jordan_Wigner,Bravyi2002,Verstraete_2005,jiang2020optimal,maskara2025fast,Gorshkov2025low}, and are consequently difficult to identify experimentally. The tricritical Ising (TCI) universality class~\cite{Blume1966,Capel1966,Friedan1984} provides a particularly intriguing setting in which this question becomes both theoretically sharp and experimentally accessible. On the theory side, it hosts an emergent spacetime supersymmetry (SUSY) \cite{Friedan1985,Zamolodchikov1986,GSV2014,Rahmani2015,Fendley2018}, which ties each emergent fermion to bosonic superpartners and thereby imposes rich relations on both the spectrum and the scaling dimensions \cite{zou2020emergence,Hsieh2021,cheng2025schwinger}. On the experimental side, several bosonic Rydberg-based models have been proposed to realize quantum version of TCI~\cite{slagle2021microscopic,Li2024,wang2025tricritical,naus2025measurement,Endres2026conformal}, making it a natural target for investigation.

What remains missing, however, is a direct probe of a single emergent fermion in a bosonic Rydberg realization of TCI. Existing evidence comes primarily from two-point correlators~\cite{Li2024}, which are ultimately bosonic observables constructed from fermion bilinears, or from indirect constraints on scaling dimensions in the bosonic sector bridged by the SUSY fermion~\cite{wang2025tricritical}. A genuine smoking-gun probe of an individual emergent SUSY fermion is therefore still lacking.

\begin{figure}[t!]
    \begin{center}
    \includegraphics[width=0.45\textwidth]{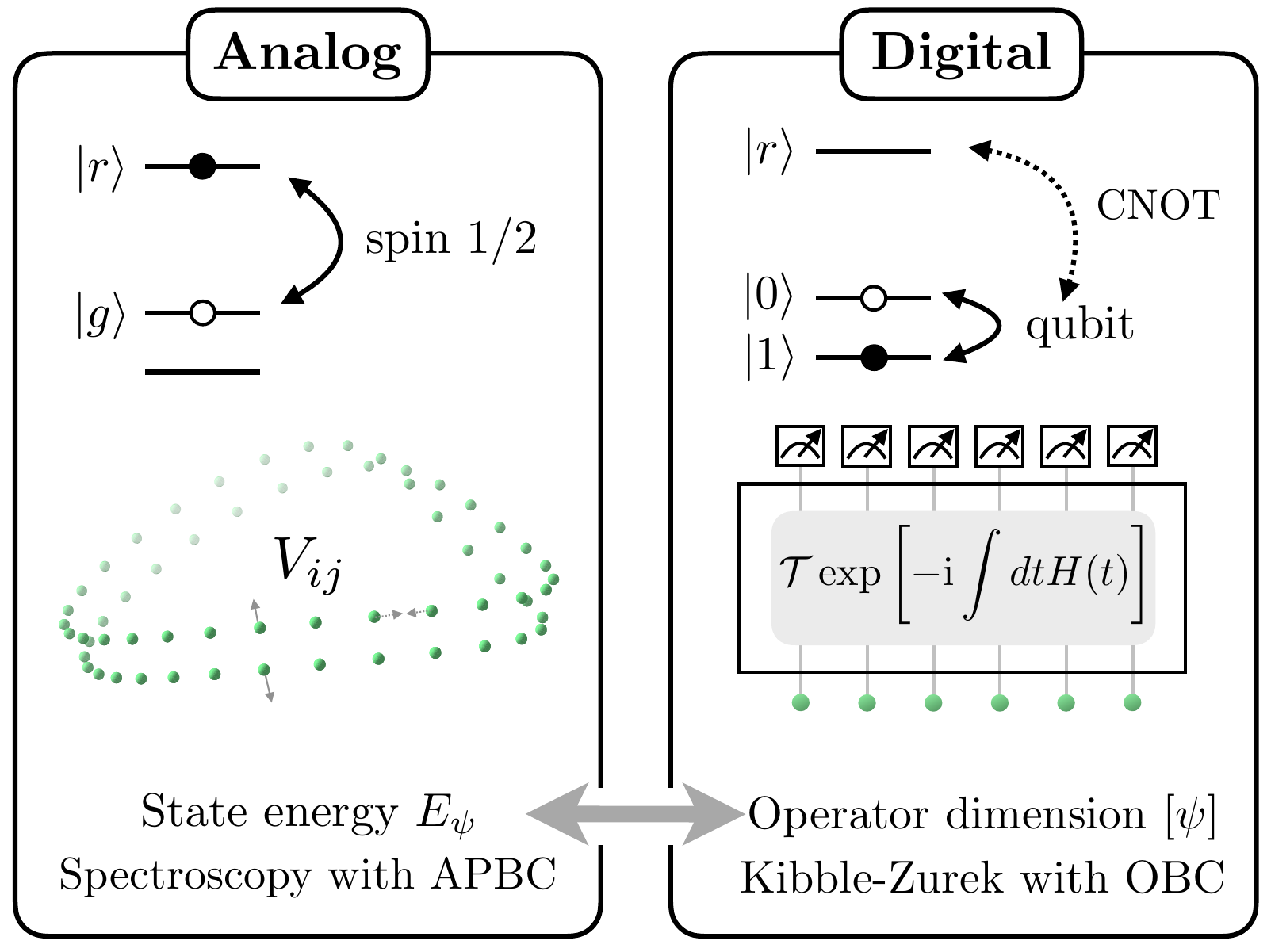}
    \caption{Schematic of two routes for probing emergent fermions with Rydberg atom simulators. (Left) In the analog mode, each atom realizes a spin-$1/2$ degree of freedom, and the geometric reconfigurability of the array, together with the interactions $V_{ij}$, constructs the tricritical Ising model with APBC. Spectroscopy provides access to the fermionic state energies $E_\psi$. (Right) In the digital mode, a qubit array with programmable gates implements the real-time evolution of a fermionic Kibble--Zurek protocol with OBC, probing the scaling dimension $[\psi]$.}
    \label{Fig:analog_digital}
    \end{center}
\end{figure}

\begin{figure}[t!]
    \begin{center}
    \includegraphics[width=0.45\textwidth]{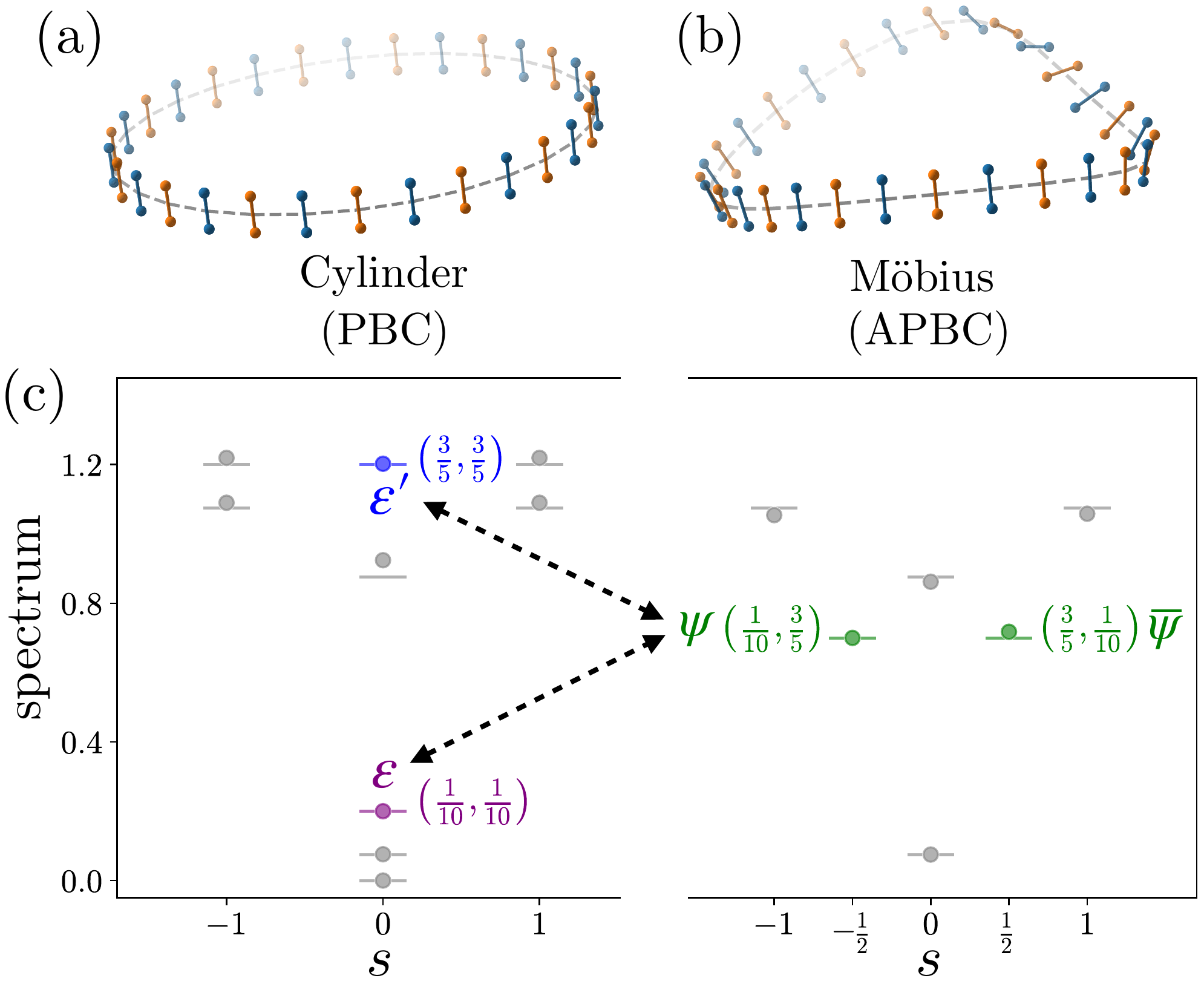}
    \caption{(a) Cylinder geometry realizing PBC with $2L$ atoms. (b) Optimized developable M\"obius geometry realizing APBC. (c) Corresponding low-energy spectra, organized by conformal spin $s$. The bosonic primaries $\varepsilon$ (blue) and $\varepsilon'$ (purple) are superpartners of the fermionic primaries $\psi$ and $\bar{\psi}$ (green).}
    \label{Fig:spectrum}
    \end{center}
\end{figure}

In this Letter, we propose two complementary routes, one analog and one digital, to unveil a single emergent fermionic mode $\psi$ through both its state and operator content, exploiting the reconfigurability of atom arrays [Fig.~\ref{Fig:analog_digital}]. In the analog mode, we realize the TCI using a dual-species Rydberg ladder, following Ref.~\cite{Li2024}. Different boundary conditions give rise to distinct spectra: under periodic boundary conditions (PBC), the lowest $\mathbb{Z}_2$-symmetric excitations are the bosonic primaries $\varepsilon$ and $\varepsilon'$, while under antiperiodic boundary conditions (APBC) the low-energy spectrum contains the fermionic mode $\psi$, the superpartner of $\varepsilon$ and $\varepsilon'$~\cite{zou2020emergence}. Accessing the state content of $\psi$ therefore requires both spectroscopic measurements and the implementation of APBC. The former is now within reach in experiments with adiabatically prepared TCI ground states on moderate system sizes~\cite{Endres2026conformal}. To achieve the latter, we employ a M\"obius geometry of the Rydberg ladder [Fig.~\ref{Fig:spectrum}(b)]. Crucially, this geometry realizes the genuine APBC spectrum, which is distinct from the spectrum of an open chain with oppositely pinned boundary fields~\cite{Endres2026conformal}.

\begin{figure*}[t!]
    \begin{center}
    \includegraphics[width=\textwidth]{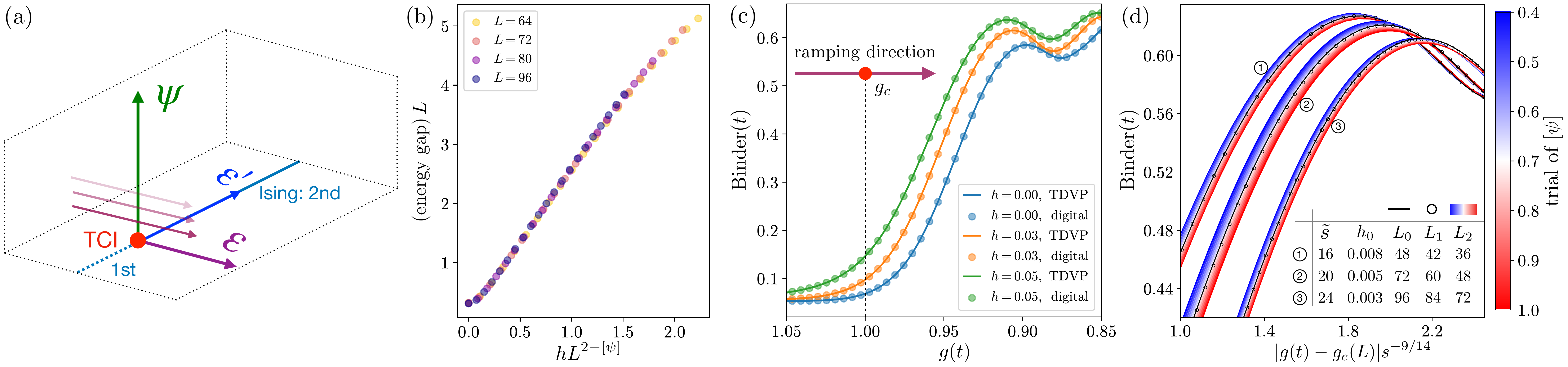}
    \caption{(a) Schematic near the TCI point in the $(\varepsilon,\varepsilon',\psi)$ parameter space, controlled by $(g-g_c,\lambda-\lambda_{\mathrm{TCI}},h)$. The pale-purple arrows indicate ramp trajectories passing near the TCI point at different fixed $h$. (b) Finite-size scaling collapse of the gap opened by $H_\chi$ at the TCI point. (c) Dynamical Binder ratio during a ramp across the transition for $L=16$, comparing TDVP lines with digital-circuit data points. (d) Fermionic Kibble--Zurek results for the three data sets labeled \textcircled{\raisebox{-0.9pt}{1}}, \textcircled{\raisebox{-0.9pt}{2}}, and \textcircled{\raisebox{-0.9pt}{3}}, each with different $\tilde{s}$ and $h_0$. Each data set contains three system sizes $(L_0,L_1,L_2)$, whose KZ curves collapse when $[\psi]=7/10$. The color scale denotes the trial value of $[\psi]$; values away from $7/10$ produce a systematic drift.}
    \label{Fig:psi_KZ}
    \end{center}
\end{figure*}

\begin{figure}[t!]
    \begin{center}
    \includegraphics[width=0.47\textwidth]{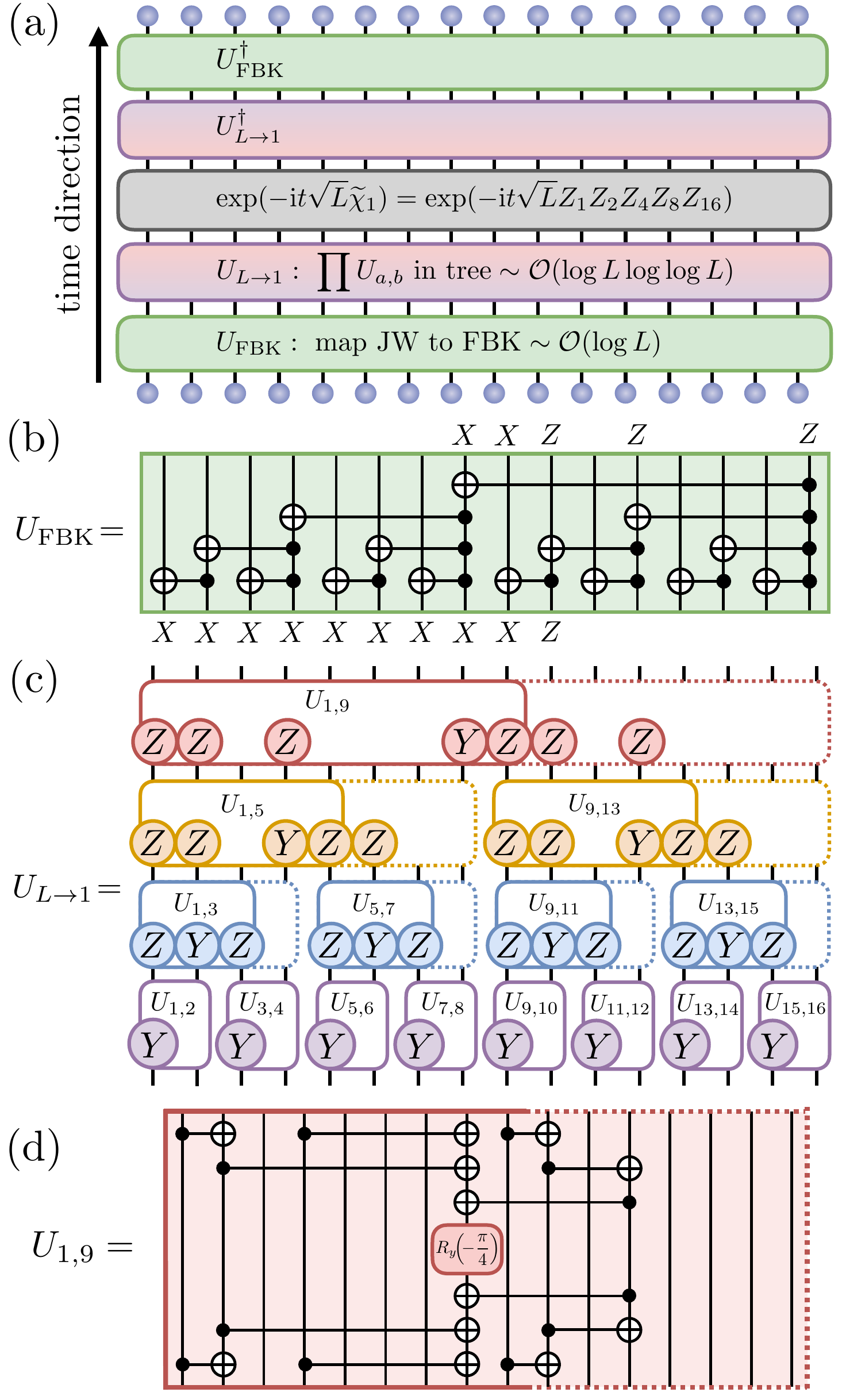}
    \caption{Circuit implementation of $e^{-\ii t H_\chi}$ in the FBK basis for $L=16$. (a) From bottom to top, $U_\fbk$ maps JW Majoranas $\chi_i$ to FBK Majoranas $\tchi_i$, $U_{L\to1}$ rotates $\sum_i\tchi_i$ to $\sqrt{L}\tchi_1$, and a single Pauli-string rotation implements time evolution $\exp(-\ii t\sqrt{L}\tchi_1)$; the inverse circuit maps back to the JW basis. (b) Long-range CNOT tree for $U_\fbk$, illustrated by $\chi_{10}=X_1\cdots X_9Z_{10}\mapsto \tchi_{10}=X_8X_9Z_{10}Z_{12}Z_{16}$. (c) Binary tree of two-Majorana rotations $U_{a,b}$ forming $U_{L\to1}$. Circled operators mark the generator $P_{a,b}$, which lies inside the dashed block $[a,2b-a-1]$ but not necessarily inside $[a,b]$. (d) Long-range CNOT decomposition of $U_{1,9}$. It has support on sites $1,2,4,8,9,10,12$, with weight $2l-1=7$ at layer $l=4$.}
    \label{Fig:circuit}
    \end{center}
\end{figure}

On the other hand, a characteristic property of an emergent particle at criticality is its scaling dimension. This operator content can be accessed in the digital mode, where we design a nonequilibrium critical protocol with a perturbation constructed from a JW Majorana string [Fig.~\ref{Fig:psi_KZ}(a)]. The resulting response obeys fermionic Kibble--Zurek scaling~\cite{kibble1976topology,zurek1985cosmological}, from which the scaling dimension $[\psi]$ can be extracted. Enabled by the reconfigurability of atom arrays, this construction can be implemented with only $\mathcal{O}(\log L \log\log L)$ overhead [Fig.~\ref{Fig:circuit}].

Together, these two routes establish the state--operator correspondence for the emergent single fermion: its spectral content and scaling dimension are two sides of the same coin in the underlying conformal field theory (CFT)~\cite{DiFrancescoCFTBook}. 
Hence, the quantitative agreement between the spectral and scaling descriptions provides a verification of conformal invariance~\cite{Rychkov2025,podo2026direct,RongRychkovToAppear} through the state--operator correspondence. As a side product, this also verifies the SUSY structure through the relations $E_\psi = E_\varepsilon + 1/2 = E_{\varepsilon'} - 1/2$ in the scaled spectrum and $[\psi] = [\varepsilon] + 1/2 = [\varepsilon'] - 1/2$ in the scaling dimensions. The detailed constructions and results for both routes are presented below.

\emph{Analog mode: M\"obius geometry and APBC.---}
We first discuss how to identify emergent fermions in the analog mode of the Rydberg platform. We begin by reviewing the Rydberg ladder model~\cite{Li2024},
\begin{equation}
H_\text{analog} = \frac{\Omega}{2} \sum_{i=1}^{2L} X_i - \Delta \sum_{i=1}^{2L} n_i + \sum_{i<j} V_{ij} n_i n_j,
\end{equation}
where we encode a two-level system within each atom with the ground and Rydberg states $\ket{g},\ket{r}$. $X=\ket{g}\bra{r}+h.c.$ couples the two levels, and $n=\ket{r}\bra{r}$ projects onto the Rydberg state. $\Omega$ is the Rabi frequency, $\Delta$ is the laser detuning, and $V_{ij}$ is the van der Waals (vdW) interaction strength between Rydberg atoms. The two atoms on each rung are arranged within the Rydberg blockade regime and only one of them can be excited. Crucially, on alternating rungs, the Rydberg states have different principal quantum numbers such that the corresponding vdW interaction is attractive~\cite{Zhan2017PRL,Zhan2022PRL,Hannes2022PRX,anand2024dual}, and the ground state goes through a $\mathbb{Z}_2$ symmetry-breaking transition from paramagnetic at small $\Delta$ to ferromagnetic at large $\Delta$, where the symmetry operation acts as an exchange of the two legs. The transition is first order at small $\Omega$ where it is more of a classical nature, and is second order at large $\Omega$ where quantum fluctuation dominates. Hence, we expect a tricritical Ising transition in between, as verified in Ref.~\cite{wang2025tricritical}.

The geometric nature of the symmetry action leads to an intriguing interplay between the atom configuration and the low-energy critical physics, where the concept of boundary condition assumes a central role. The significance of boundary conditions in critical systems is that it enforces symmetry constraints on possible low-energy states and, in our case, allows a direct addressing of the fermionic states. Focusing on the $\mathbb{Z}_2$ case, two natural choices are the PBC and the APBC. They are defined by whether a $\mathbb{Z}_2$ defect is inserted as one travels along the periodically wrapped system once. Translated into the geometric language at hand, this means whether the atoms are topologically assembled as a cylinder or a M\"obius band. 

Notably, such a topologically nontrivial configuration has been experimentally demonstrated as a tour de force of the tweezer array technique, alongside an atom array miniature of the Eiffel Tower \cite{Barredo2018}. However, closer inspection reveals that what has been assembled is a ``planar'' M\"obius band, where the centerline of the band lies on a plane. This leads to significant distortion of the ladder as the band twists in space and explicitly breaks the $\mathbb{Z}_2$ symmetry, destroying the phase transition altogether.

To fix this issue, we borrow ideas of the ``developable'' M\"obius band~\cite{Mobius2007}, a topic of interest in materials science where how materials bend and twist in a smoothest way is a natural concern. In particular, researchers in this field have discussed what shapes M\"obius bands naturally assume when the underlying material is unstretchable, hence the term developable~\footnote{The mathematical definition of a developable surface is that the Gaussian curvature is zero.}. The general conclusion is that the centerline of the band must not be planar, as one intuitively expects from real-life experience with paper belts.

With this materials science notion in mind, we design a loss function that minimizes distortions and hence deviation from an ideal APBC Hamiltonian; see End Matter for more details. The resulting configuration and the corresponding scaled low-energy spectrum obtained via density-matrix renormalization group (DMRG) \cite{white1992density} are shown in Fig.~\ref{Fig:spectrum}, where we also include the PBC result for comparison.

CFT provides precise predictions of the energy and operator correspondence of the low-energy states. The energy levels are given by $E_0+2\pi v/L(h_L+h_R+n)$, where $h_L,h_R$ are the chiral conformal dimension of the primary fields and $n$ is the level. In particular, the first excited states in the APBC sector map to the fermionic operator, with $(h_L,h_R)=(1/10,3/5),(3/5,1/10)$ \cite{Friedan1985,DiFrancescoCFTBook}. By performing spectroscopic measurements~\cite{Endres2026conformal} in the APBC sector, these numbers can be measured which uniquely identifies their fermionic nature, and the scaling dimension of the emergent fermion $\psi$, namely $[\psi]=7/10$.

\emph{Digital mode.---}
We now turn to the digital mode and ask how to probe the scaling dimension of $\psi$ directly. Since the underlying hardware is bosonic, accessing emergent fermionic degrees of freedom requires nonlocal operators. This makes the digital mode, where JW strings can be compiled explicitly, more natural than a purely analog implementation based only on nearby interactions.

Unlike the Rydberg blockade model used in the analog protocol, it is convenient here to work with a lattice Hamiltonian that is more suited to gate-based implementation of a supersymmetric tricritical point. We consider the O'Brien--Fendley (OF) Hamiltonian~\cite{Fendley2018,slagle2021microscopic},
\begin{equation}
H_{\mathrm{OF}} = H_{ZZ} + g H_X + \lambda H_3,
\end{equation}
\begin{align} 
H_{ZZ} = -\sum_{i} Z_{i} Z_{i+1},\,
H_X = - \sum_{i} X_{i},\\ 
H_3 = \sum_{i}( X_{i-1} Z_{i} Z_{i+1} + Z_{i-1} Z_{i} X_{i+1}).
\end{align}
Tricriticality occurs at $g=g_c=1$ and $\lambda=\lambda_{\mathrm{TCI}}\approx 0.428$. In the continuum description, the deviation $g-g_c$ couples to $\varepsilon$ and tunes the system across the disordered-ordered transition, while $\lambda-\lambda_{\mathrm{TCI}}$ couples to $\varepsilon'$ and drives the system away from tricriticality. Accordingly, within the $(\varepsilon,\varepsilon')$ plane the OF model exhibits the standard near-TCI structure [Fig.~\ref{Fig:psi_KZ}(a)]: for $\lambda>\lambda_{\mathrm{TCI}}$ the transition tuned by $g$ lies in the Ising universality class, while for $\lambda<\lambda_{\mathrm{TCI}}$ it becomes first order.

In this model, the relevant fermionic degrees of freedom can be represented by JW Majorana operators,
\begin{equation}
\chi_i=\left(\prod_{l=k}^{i-1}X_l\right)Z_i,
\label{eq:JW}
\end{equation}
which satisfy $\{\chi_i,\chi_j\}=2\delta_{ij}$. Previous work has focused primarily on computing two-point functions such as $\langle \chi_i \chi_j\rangle$ in eigenstates or thermal states of the Hamiltonian \cite{Fendley2018,Li2024}. Here we adopt a different viewpoint, closer in spirit to perturbation--response theory: we add the Majorana operator directly to the Hamiltonian, $H_\text{digital} = H_{\mathrm{OF}} + h H_\chi$ with $H_\chi = \sum_i \chi_i$, and then ask how the system responds. 

At first sight, a term linear in $\chi_i$ may seem unnatural, since it breaks fermion parity. In the present setting, however, the microscopic simulator is bosonic, so there is no fundamental obstruction to implementing such a perturbation. The real question is whether the response to this perturbation encodes the scaling dimension $[\psi]$. The answer hinges on the boundary conditions. While the analog approach probes the spectrum under closed boundary conditions, our digital protocol is naturally formulated with open boundary conditions (OBC), where the JW string can terminate topologically on a symmetry-preserving boundary. This boundary endpoint carries vanishing scaling dimension and therefore avoids dressing the bulk endpoint of the string by additional scaling fields, allowing the perturbation to isolate $[\psi]$; see End Matter for more details. As a result, perturbing by $H_\chi$ under OBC opens a gap $\delta E \sim h^{\nu_\psi}$ near TCI, with $\nu_\psi^{-1} = 2-[\psi]$. For a finite system, this implies the scaling form $\delta E \sim \frac{1}{L} f(h L^{2-[\psi]})$, which we confirm by DMRG in Fig.~\ref{Fig:psi_KZ}(b).

\emph{Fermionic Kibble--Zurek.---}
The gap opened by $H_\chi$ at the TCI point can be probed dynamically: by varying the ramp speed near the phase transition, the universal scaling associated with this gap becomes directly accessible through the Kibble--Zurek (KZ) mechanism~\cite{kibble1976topology,zurek1985cosmological}. In the absence of the fermionic perturbation ($h=0$), i.e., within the $(\varepsilon,\varepsilon')$ plane in Fig.~\ref{Fig:psi_KZ}(a), a near-TCI KZ protocol has been proposed for analog Rydberg simulators that exhibits bosonic KZ scaling for sufficiently slow ramps, from which the scaling dimensions of $\varepsilon$ and $\varepsilon'$ can be extracted robustly~\cite{wang2025tricritical,yin2025driven}.

Building on this idea, we now add a weak fermionic perturbation of fixed strength $h$, corresponding to shift along the $\psi$ direction in Fig.~\ref{Fig:psi_KZ}(a), and ramp from the disordered phase to the ordered phase by tuning the transverse-field strength $g$. More explicitly, we fix $H_0 = H_{ZZ} + \lambda H_3 + h H_\chi$, and evolve with
\begin{equation}
\mathcal{U}(t) = \mathcal{T}\exp\left[-\ii \int_0^{t} dt' \Big(H_0 + g(t') H_X\Big)\right].
\end{equation}
Starting from the ground state $|\text{GS}\rangle$ deep in the disordered regime, we then evolve the wavefunction $|\Psi(t)\rangle = \mathcal{U}(t)|\mathrm{GS}\rangle$ and monitor the Binder ratio, 
\begin{equation}
\text{Binder}(t) = 1 - \frac{1}{3}
\frac{\langle \Psi(t)|M^4|\Psi(t)\rangle}{\langle \Psi(t)|M^2|\Psi(t)\rangle^2},
\end{equation}
where $M=\sum_i Z_i$ is the order parameter operator. We propose the universal KZ scaling form
\begin{equation}
\text{Binder}(s,g,h,L) = \mathcal{F}\left(sL^{3-[\varepsilon]},\,(g-g_c)L^{2-[\varepsilon]},\,hL^{2-[\psi]}\right).
\end{equation}
The first two arguments are the standard bosonic finite-size KZ scaling variables \cite{Grandi2011,Kolodrubetz2012,Chandran2012,Huang2014FTS,rossini2021coherent}, associated respectively with the finite ramp speed $s$ and the detuning $g-g_c$ from criticality, while the third argument encodes the additional relevant fermionic perturbation induced by $H_\chi$.

We test this scaling law using the time-dependent variational principle (TDVP) \cite{TDVP}. Figure~\ref{Fig:psi_KZ}(c) shows the dynamical Binder ratio as a function of $g(t)$ for $L=16$ and several values of $h$. To demonstrate the fermionic scaling curve directly, we fix the scaled ramp rate $\tilde{s}=sL^{3-[\varepsilon]}$ \cite{zhang2025observation} and choose three system sizes $(L_0,L_1,L_2)$ together with couplings $(h_0,h_1,h_2)$ such that $h_0L_0^{2-[\psi]} = h_1L_1^{2-[\psi]} = h_2L_2^{2-[\psi]}$. When the Binder ratio is plotted against $|g(t)-g_c(L)|s^{-(2-[\varepsilon])/(3-[\varepsilon])}$, the data collapse is optimized at $[\psi]=7/10$, as shown in Fig.~\ref{Fig:psi_KZ}(d). Choosing a trial value of $[\psi]$ away from $7/10$ produces a systematic drift and spoils the collapse.

\emph{Shallow-circuit implementation of JW strings via reconfigurability.---}
Simulating fermions on bosonic hardware typically incurs substantial overhead~\cite{Bravyi2002}. The difficulty is transparent here because the perturbation $H_{\chi}$ is highly nonlocal in the computational basis. A direct implementation of $e^{-\ii t H_{\chi}}$ would build up the JW strings site by site, yielding a circuit depth that grows linearly with the system size $L$~\cite{kokcu2024linear}. In a tweezer-based system, however, atoms can be rearranged between entangling layers, so that effectively long-range two-qubit gates are available~\cite{Bluvstein2022,bluvstein2024logical,evered2026nonlocal}. 
This reconfigurability enables a much shallower compilation based on binary-tree compression. In particular, a binary tree of two-Majorana rotations reduces the sum of Majoranas to a single Majorana in depth $\log L$. Each two-Majorana rotation is a Pauli-string rotation of weight at most $L$, which can be implemented by an additional CNOT tree in depth $\log L$. This gives an overall depth $\mathcal{O}((\log L)^2)$ for the direct JW construction.

The depth can be reduced further by combining nonlocal two-qubit gates with the Fenwick--Bravyi--Kitaev (FBK) mapping~\cite{Fenwick1994,Bravyi2002}. Intuitively, this map logarithmically shrinks the weight of the fermionic operators, i.e., the length of the long JW strings. We prove that the FBK compression remains compatible with the binary-tree compression introduced above, so that the two can be applied in tandem. Together they compress the final circuit depth to $\mathcal{O}(\log L\,\log\log L)$. The full circuit for $L=16$ is shown in Fig.~\ref{Fig:circuit} and generalizes straightforwardly to other $L$, with detailed design principle and proofs deferred to the End Matter.

We use this gate-level construction to simulate the real-time evolution with the time-evolving block decimation (TEBD) \cite{TEBD} method. The resulting circuit data (dots) agree with TDVP time evolution obtained directly from the Hamiltonian (lines), as shown in Fig.~\ref{Fig:psi_KZ}(c).

\emph{Outlook.---}
We have proposed two complementary protocols, one analog and one digital, that together exploit the geometric reconfigurability of Rydberg atom simulators to unveil emergent fermions in SUSY critical quantum matter. The central feature enabling both protocols is the movable-atom architecture, which opens several promising directions for further investigation. On the analog side, the developable M\"obius band is only one entry in a much richer family of geometries accessible in three dimensions \cite{Rui2025}; more elaborate constructions could realize a variety of nontrivial boundary and defect conditions \cite{wang2025kink,Sarma2026,Rong2026interface}, and thereby probe topological sectors and symmetry-enriched criticality beyond the APBC case studied here. On the digital side, our perturbation--response strategy is not tied to TCI: any critical point with JW-encoded  emergent fermions, or more generally parafermions~\cite{Dotsenko1984,Nienhuis1985,Mong2014}, should be accessible by the same scheme. Finally, the nonlocal gates made possible by atom rearrangement raise new questions of their own: at realistic gate fidelities, the resulting noise acts as a nonlocal, string-like form of disorder \cite{evered2026nonlocal}, qualitatively distinct from the local perturbations considered in most previous studies of disordered critical systems \cite{li2025random,Chepiga2026Randomness}. Understanding its universal effects on critical dynamics, entanglement, and operator scaling is a compelling direction in its own right, and one that reconfigurable neutral-atom platforms are uniquely positioned to explore.

\emph{Acknowledgements.---}
We are grateful to Pengfei Zhang, Junchen Rong, and Slava Rychkov for useful discussion. This work is supported by Quantum Science and Technology-National Science and Technology Major Project under Grant No.~2025ZD0300400 and National Natural Science Foundation of China under Grant No.~12504307 (C.L.). C.L. is also supported by Tsinghua University Dushi program. H.W. is supported by China Postdoctoral Science Foundation under Grant No.~2024M751609 and Postdoctoral Fellowship Program of CPSF under Grant No.~GZC20231364. S.L. acknowledges support from the Chinese Academy of Sciences (CAS) under Grant No. YSBR-150 and a startup fund from the Institute of Physics, CAS.
The DMRG, TDVP and TEBD calculations are performed using the ITensor library \cite{ITensor}.

\bibliography{unleash_fermion.bbl}

\newpage
\begin{center}
{\large\bfseries End Matter}
\end{center}
\setlength{\abovedisplayskip}{4pt plus 1pt minus 2pt}
\setlength{\belowdisplayskip}{4pt plus 1pt minus 2pt}
\setlength{\abovedisplayshortskip}{2pt plus 1pt minus 1pt}
\setlength{\belowdisplayshortskip}{3pt plus 1pt minus 1pt}
\setlength{\textfloatsep}{6pt plus 1pt minus 2pt}
\setlength{\floatsep}{6pt plus 1pt minus 2pt}

\setcounter{equation}{0}
\renewcommand{\theequation}{A\arabic{equation}}
\renewcommand{\theHequation}{A\arabic{equation}}
\emph{Appendix A: Optimization of the developable M\"obius band.---}
For our purposes, we wish the $\mathbb{Z}_2$ and translation symmetries to be as intact as possible, and we perform a numerical optimization of the M\"obius band configuration to achieve our goal. We define a loss function
\begin{equation}
\begin{split}
&\mathcal{L}(\{\br_i\})=\sum_{i}(|\br_{i}-\br_{i+1}|-r_1)^2\\[-0.5em]
&+(|\br_{i}-\br_{i+L+1}|-|\br_{i+1}-\br_{i+L}|)^2\\
&+\alpha(|\br_{i}-\br_{i+L+1}|-|\br_{i+1}-\br_{i+L+2}|)^2\\
&+\beta(|\br_{i}-\br_{i+L}|-r_2)^2/2+\beta(|\br_{i}-\br_{i+L+1}|-r_3)^2\\
&+\gamma(|\br_{i}-\br_{i+2}|-2r_1)^2+\gamma(|\br_{i}-\br_{i+L+2}|-r_4)^2,    
\end{split}
\end{equation}
where $r_1$ and $r_2$ are targeted lattice constants along the leg and rung directions, $i=1,\dots,2L$ periodically traverses the sites such that $i$ and $i+L$ belong to the same rung, $r_3=\sqrt{r_1^2+r_2^2}$, $r_4=\sqrt{4r_1^2+r_2^2}$, and $\alpha,\beta,\gamma$ are hyperparameters that prioritize translational and $\mathbb{Z}_2$ symmetries with the targeted lattice constants as a reference~\footnote{In the calculation we use $\alpha=2.5,\beta=0.005,\gamma=0.0025$.}. The loss function can be viewed as the classical potential energy by coupling atoms with springs of designated rest length.

\setcounter{equation}{0}
\renewcommand{\theequation}{B\arabic{equation}}
\renewcommand{\theHequation}{B\arabic{equation}}
\emph{Appendix B: CFT boundary conditions: spectrum and scaling dimensions.---}
At the critical point, boundary conditions can strongly affect both the finite-size spectrum and the scaling behavior of nonlocal operators. We first discuss the spectrum. For closed chains, PBC and APBC select distinct sectors of the TCI CFT. As discussed in the main text, the PBC spectrum contains the lowest $\mathbb{Z}_2$-symmetric bosonic primaries $\varepsilon$ and $\varepsilon'$, while the APBC spectrum contains the fermionic sector, including the mode $\psi$. One may wonder whether an open chain with oppositely pinned boundary fields realizes the same low-energy physics as closed APBC. Although such boundary pinning is APBC-like in the sense that it enforces a domain wall, its finite-size spectrum is that of a boundary CFT on an interval. For opposite fixed-type TCI boundaries, the allowed states form the chiral $\varepsilon''$ boundary tower~\cite{Endres2026conformal}. This is distinct from the full left--right operator content of the closed-chain APBC sector, which contains the fermionic fields $\varepsilon'_L\varepsilon_R$ and $\varepsilon'_R\varepsilon_L$ with scaling dimension $7/10$. Thus, an oppositely pinned open chain does not realize the genuine APBC spectrum on a circle, unlike the M\"obius geometry used in the main text.

We next discuss the scaling dimension extracted from a JW string. In a closed boundary case, the starting point of the JW string ($k$ in Eq.~\eqref{eq:JW}) carries an additional nontrivial scaling dimension. In field-theory language, the JW string is a topological defect line \cite{Yin2019TDL} generating the $\mathbb{Z}_2$ symmetry of the theory. The $\mathbb{Z}_2$ odd endpoint of the line (the one with a $Z$ insertion) corresponds to the fermion operator of interest, while the other $\mathbb{Z}_2$ even endpoint corresponds to the disorder operator $\mu$. Schematically, the continuum operator is therefore $\chi_{\mathrm{PBC}}(x_0-x) \sim \mu(x_0)\psi(x)$, so the target scaling dimension is dressed by the disorder field and is not singled out. OBC is qualitatively different, because the JW string can now have a boundary endpoint of vanishing scaling dimension. As mentioned previously, the JW string is a topological defect, which means that the string can be continuously deformed in the two-dimensional Euclidean spacetime while preserving all physical quantities \cite{Yin2019TDL}. This deformability is a consequence of the bulk $\mathbb{Z}_2$ symmetry. Now suppose we take the OBC and that the $\mathbb{Z}_2$ symmetry is preserved by the boundary as well, it is then natural to expect that the JW string can have a topological (deformable) endpoint on the boundary. Such a topological endpoint will necessarily have zero scaling dimension, because its correlation function with another boundary defect endpoint operator is independent of their separation.

\begin{figure*}[t!]
    \begin{center}
    \includegraphics[width=0.75\textwidth]{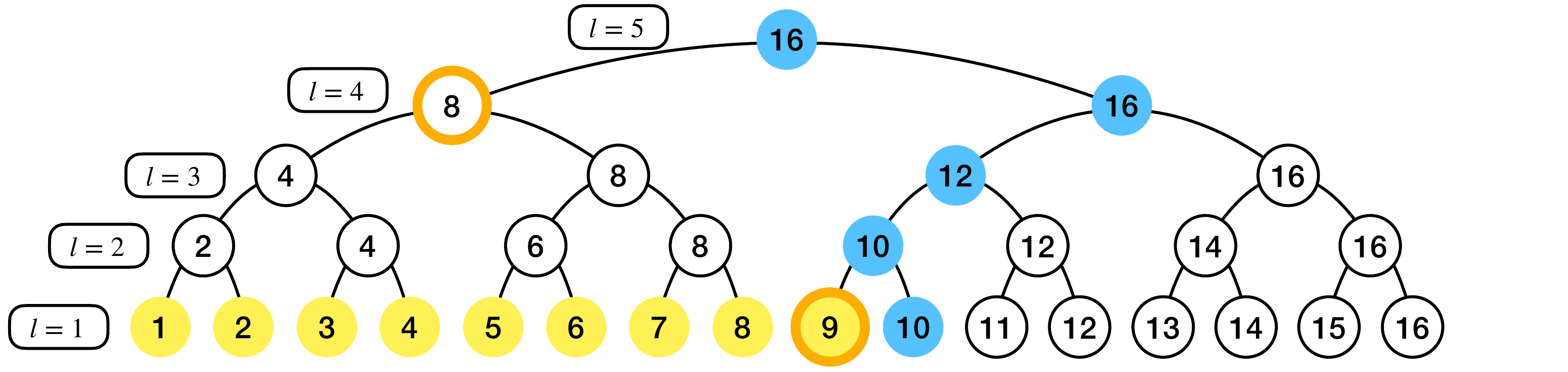}
    \caption{The Fenwick tree for $L=16$ with five layers, $l=1,...,5$. In each layer other than $l=1$, a node has the same number as its right-hand-side leaf. Each node by definition ``controls'' all other nodes below. With the Fenwick tree, one can easily transform the JW-related operators $X_1\cdots X_i$ and $Z_i$ into the FBK basis. Taking the same example as in Fig.~\ref{Fig:circuit}(b), $X_1\cdots X_9$ is transformed to $X_8X_9$ because the nodes 8 and 9 (circled in orange) control the lowermost leaves 1 through 9 (highlighted in yellow), with $\Lambda(9)=\{8,9\}$. On the other hand, $Z_{10}$ is transformed to $Z_{10}Z_{12}Z_{16}$ (highlighted in blue), by tracking which nodes are traversed when one starts from 10 and goes all the way up.  }
    \label{fig:fenwick}
    \end{center}
\end{figure*}

\setcounter{equation}{0}
\renewcommand{\theequation}{C\arabic{equation}}
\renewcommand{\theHequation}{C\arabic{equation}}
\emph{Appendix C: Shallow circuit design.---}
In this section, we provide a high-level description of the circuit design, and relegate further details to the next section. The Fenwick tree is a popular data structure that enables efficient cumulative sum calculations. Bravyi and Kitaev proposed how this construction can be used to compress the weight of JW strings, and hence Majorana operators, to $\mathcal{O}(\log L)$. As shown in the next section, this basis change is implemented by the CNOT binary-tree unitary $U_\fbk$ in Fig.~\ref{Fig:circuit}(b), with circuit-depth $\mathcal{O}(\log L)$. 

After going to the FBK basis, which maps $\chi_a \mapsto \tchi_a$, we now need to implement $\exp(-\ii t H_\chi)\mapsto\exp(-\ii t \sum_a\tchi_a)$. This is carried out by introducing a two-Majorana gate $U_{a,b}= \exp\left(\frac{\pi}{8}\tchi_a\tchi_b\right)$, which rotates $\tchi_a+\tchi_b$ into $U_{a,b}(\tchi_a+\tchi_b) U_{a,b}^{\dagger}\propto \tchi_a$, see Fig.~\ref{Fig:circuit}(c). A binary-tree compilation of these gates $U_{L\rightarrow1}$ then brings $\exp(-\ii t \sum_a\tchi_a)$ to $\exp(-\ii t\sqrt L \tchi_1)$.
Note that for the tree to be well defined, we need to ensure that $U_{a,b}$ within each layer has non-overlapping supports, which is shown in the next section.

What remains is the circuit implementation of the one- and two-Majorana gates $\exp(-\ii t \sqrt L\tchi_1)$ and $U_{a,b}$. Both turn out to take the form of a Pauli-string exponential $\exp(\ii \theta P)$, where $\theta$ is real and $P$ contains at most one $Y$ and otherwise only $Z$'s  (see Fig.~\ref{Fig:circuit}(c) and the next section).
Using $\mathrm{CNOT}_{i \to j}\, Z_iZ_j\, \mathrm{CNOT}_{i \to j}=Z_j$, a binary-tree of CNOT gates maps $\exp(\ii \theta P)$ to a single-qubit unitary, 
so its depth is logarithmic in the weight of $P$.
A representative case of $U_{1,9}$ is shown in Fig.~\ref{Fig:circuit}(d).

We now count the total circuit depth of our protocol. As shown in the next section, within $U_{L\rightarrow1}$, each $U_{a,b}$ in the $l$-th layer involves a Pauli string of weight $\mathcal{O}(l)$ (rather than the general FBK weight $\mathcal{O}(\log L)$), and therefore requires a circuit of depth only $\mathcal{O}(\log l)$. Summing over the $\mathcal{O}(\log L)$ layers, the total depth of $U_{L\rightarrow1}$ is of order $\sum_{l=1}^{\log L}\log l \sim \log L \log\log L$.

\setcounter{equation}{0}
\renewcommand{\theequation}{D\arabic{equation}}
\renewcommand{\theHequation}{D\arabic{equation}}
\emph{Appendix D: FBK basis change.---}
The FBK mapping is best understood through the Fenwick tree, as shown in Fig.~\ref{fig:fenwick}. The Fenwick tree is constructed from lower to upper, with nodes in the lowest layer labeled 1, ..., $L$, where $L$ is the system size. In each upper layer, the node is labeled by the same number as its right-hand-side leaf. The labels on the tree then defines a ``controlling''  relation among numbers, with each node controlling all the nodes below (including itself). 
For example, in Fig.~\ref{fig:fenwick}, the orange-circled node labeled 8 controls the leaves $1,\ldots,8$, highlighted in yellow, while the orange-circled node labeled 9 controls only the yellow-highlighted leaf 9, namely itself.

The Fenwick tree has a highly similar structure to the unitary $U_\fbk$. Indeed, one can readily finds out $U_\fbk X_1\cdots X_i U_\fbk^\dg=\prod_{j\in \Lambda(i)}X_j$ and $U_\fbk Z_i U_\fbk^\dg=\prod_{j\in R(i)}Z_j$ by using the Fenwick tree, together with the properties, 
\begin{equation}
\begin{split}
&\mathrm{CNOT}_{i \to j}\, Z_iZ_j\, \mathrm{CNOT}_{i \to j}=Z_j,\\
&\mathrm{CNOT}_{i \to j}\, Z_i\, \mathrm{CNOT}_{i \to j}=Z_i,\\
&\mathrm{CNOT}_{i \to j}\, X_iX_j\, \mathrm{CNOT}_{i \to j}=X_i,\\
&\mathrm{CNOT}_{i \to j}\, X_j\, \mathrm{CNOT}_{i \to j}=X_j,\\
\end{split}
\end{equation}
or graphically,
\begin{equation}
\begin{split}
&\begin{quantikz}[row sep={0.7cm,between origins}]
\lstick{} & \gate{Z} & \targ{} & \gate[2,style={draw=none}]{=} &  \targ{} & \gate{Z} & \rstick{} \\
\lstick{} & \qw & \ctrl{-1} & & \ctrl{-1}  & \gate{Z} & \rstick{}
\end{quantikz},\\
&\begin{quantikz}[row sep={0.7cm,between origins}]
\lstick{} & \qw & \targ{} & \gate[2,style={draw=none}]{=} & \targ{} & \qw & \rstick{} \\
\lstick{} & \gate{Z} & \ctrl{-1} & & \ctrl{-1}  & \gate{Z} & \rstick{}
\end{quantikz},\\
&\begin{quantikz}[row sep={0.7cm,between origins}]
\lstick{} & \gate{X} & \targ{} & \gate[2,style={draw=none}]{=} & \targ{} & \qw & \rstick{} \\
\lstick{} & \gate{X} & \ctrl{-1} & & \ctrl{-1}  & \gate{X} & \rstick{}
\end{quantikz},\\
&\begin{quantikz}[row sep={0.7cm,between origins}]
\lstick{} & \gate{X} & \targ{} & \gate[2,style={draw=none}]{=} & \targ{} & \gate{X} & \rstick{} \\
\lstick{} & \qw & \ctrl{-1} & & \ctrl{-1} & \qw & \rstick{}
\end{quantikz}.
\end{split}
\end{equation}
The set $\Lambda(i)$ consists of the uppermost nodes that precisely control the lowermost leaves 1, ..., $i$ once. Hence, continuing from the example from above, $\Lambda(8)=\{8\}$, and $\Lambda(9)=\{8,9\}$. On the other hand, the set $R(i)$ contains all nodes upward from $i$ until the uppermost is reached. Thus, $R(10)=\{10,12,16\}$, as indicated by the blue-highlighted nodes in Fig.~\ref{fig:fenwick}. It is easy to see that $|\Lambda(i)|$ and $|R(i)|$ are both bounded by $\mathcal{O}(\log L)$, hence the weight of Majorana operators in the FBK basis.

We are now ready to prove the claims regarding $U_{a,b}$ in the previous section. $U_{a,b}$ involves $\tchi_a$ and $\tchi_b$, and therefore we need to multiply $\prod_{j\in \Lambda(a-1)}X_j$, $\prod_{j\in \Lambda(b-1)}X_j$, $\prod_{j\in R(a)}Z_j$, and $\prod_{j\in R(b)}Z_j$, up to a coefficient taken care by the one-Majorana gate. Crucially, in the circuit $U_{L\rightarrow1}$, only very special $U_{a,b}$ appear: in the $l$-th layer of the circuit, $a,b$ are the two left-most leaves from a particular node $c$ on the $(l+1)$-th layer in the Fenwick tree. For example, going down from node $c=8$ on the fourth layer, we find $(a,b)=(1,5)$, and $(a,b)=(5,7)$ from node $c=8$ on the third layer. Explicitly, for $L=2^r$, the pairs are given by $(a=1+2^ls,b=a+2^{l-1})$, with $s=0,1,...,2^{r-l}-1$, as shown by the solid boxes for all $(a,b)$ in Fig.~\ref{Fig:circuit}(c). 

For these particular choices of $(a,b)$, $\Lambda(b-1)$ is the union of $\Lambda(a-1)$ and $\{b-1\}$, hence the product of the two $X$ strings is just $X_{b-1}$. On the other hand, $c=a+2^l-1$ is common controlling node of $(a,b)$, then $R(a)=\{a,a+2^1-1,a+2^2-1,...,a+2^{l-2}-1,a+2^{l-1}-1=b-1\}\cup R(c)$ and $R(b)=\{b,b+2^1-1,b+2^2-1,...,b+2^{l-2}-1\}\cup R(c)$. The shared part $R(c)$ cancels in the product, leaving the $Z$ string $Z_aZ_{a+2^1-1}\cdots Z_{a+2^{l-2}-1}Z_{b-1}Z_{b}Z_{b+2^1-1}\cdots Z_{b+2^{l-2}-1}$. Combining the $X$ and $Z$ contributions, the Pauli string associated with $U_{a,b}$ is
\begin{equation}
\underbrace{Z_aZ_{a+2^1-1}\cdots Z_{a+2^{l-2}-1}}_{\substack{l-1}}
\,Y_{b-1}\,
\underbrace{Z_bZ_{b+2^1-1}\cdots Z_{b+2^{l-2}-1}}_{\substack{l-1}} .
\end{equation}
Thus $U_{a,b}$ has weight $2l-1$ and support within $[a,c=2b-a-1]$, the dashed box in Fig.~\ref{Fig:circuit}(c). Since the supports within each layer are non-overlapping, the binary tree $U_{L\rightarrow1}$ is well defined, and the weight $2l-1$ implies a CNOT-tree depth of only $\mathcal{O}(\log l)$, reproducing the count of the previous section.

\end{document}